\documentclass{iopjournal}

\begin{document}

\articletype{Article type} 

\title{Quantum Pontus--Mpemba Effect Enabled by the Liouvillian Skin Effect}

\author{Stefano Longhi}

\affil{$^1$Dipartimento di Fisica, Politecnico di Milano, Piazza L. da Vinci 32, I-20133 Milano, Italy}

\affil{$^2$IFISC (UIB-CSIC), Instituto de F\'isica Interdisciplinar y Sistemas Complejos, Campus Universitat de les Illes Balears, E-07122 Palma de Mallorca, Spain}

\email{stefano.longhi@polimi.it}


\begin{abstract}
We unveil a {quantum Pontus--Mpemba effect} enabled by the {Liouvillian skin effect} in a dissipative tight-binding chain with asymmetric incoherent hopping and coherent boundary coupling. The skin effect, induced by non-reciprocal dissipation, localizes relaxation modes near the system boundaries and gives rise to non-orthogonal spectral geometry. While such non-normality is often linked to slow relaxation, we show that it can instead \emph{accelerate} relaxation through a two-step protocol -- realizing a quantum Pontus--Mpemba effect.
Specifically, we consider a one-dimensional open chain with coherent hopping $J$, asymmetric incoherent hoppings $J_{\rm R} \neq J_{\rm L}$, and a controllable end-to-end coupling $\epsilon$. For $\epsilon=0$, the system exhibits the Liouvillian skin effect, with left and right eigenmodes localized at opposite edges. We compare two relaxation protocols toward the same stationary state: (i) a direct relaxation with $\epsilon=0$, and (ii) a two-step (Pontus) protocol where a brief coherent evolution transfers the excitation across the lattice before relaxation. Although both share the same asymptotic decay rate, the two-step protocol relaxes significantly faster due to its reduced overlap with the slow boundary-localized Liouvillian mode. The effect disappears when $J_{\rm R}=J_{\rm L}$, i.e., when the skin effect vanishes. 
Our results reveal a clear connection between boundary-induced non-normality and protocol-dependent relaxation acceleration, suggesting new routes for controlling dissipation and transient dynamics in open quantum systems.
\end{abstract}

\section{Introduction}
The Mpemba effect ~\cite{r1,r2,r3}, in which initially hotter systems can relax faster than colder ones, has become a widely studied topic in nonequilibrium physics \cite{r4}. This behavior has been observed and analyzed in a variety of classical systems \cite{r4,r5,r6,r7,r8,r9,r9b,r9c,r10,r11,r12}, including granular fluids~\cite{r5}, supercooled liquids~\cite{r6}, colloidal suspensions~\cite{r8,r10}, and spin glasses~\cite{r9,r9b,r9c}, among others (see the recent overview~\cite{r12}). Proposed explanations range from system-specific effects, such as evaporation and convection in fluids, to more general mechanisms associated with nontrivial overlap with slow modes in Markovian relaxation dynamics and kinetic-state space structure~\cite{r4,r6,r7,r11}.
The conceptual leap to quantum systems led to the notion of a \emph{quantum Mpemba effect} (QME) \cite{r12,r12b}, which has been explored in both open 
\cite{r13,r14,r15,r16,r17,r18,r19,r20,r21,r22,r28,r29,r29b,r30,r31,r34,r35,r35b,r36,r37,r38,r40,r42,r43,r44,r45,r45b,r50,r51,r52,r52b,r53,r54} and closed \cite{r18a,r23,r24,r25,r32,r33,r26,r27,r39,r41} quantum dynamics; for recent reviews see e.g. \cite{r12,r12b}. Studies have shown QME in open Markovian systems governed by Lindblad dynamics~\cite{r13,r14,r16,r28,r30,r34,r36,r45b}, in closed many-body and integrable dynamics~\cite{r15,r23,r24,r25}, and in non-Hermitian or effectively non-unitary setups~\cite{r17,r19,r20,r22,r31}. Several works clarified mechanisms by which certain initial quantum states can relax faster due to smaller projection on slow decaying Liouvillian modes, non-normal dynamics, or due to dynamical symmetry restoration and exceptional point physics~\cite{r14,r19,r21,r22,r29}. Reviews and recent surveys summarize these developments~\cite{r12b,r12}.
Very recently, a further generalization named the \emph{Pontus--Mpemba effect} (PME) was put forward \cite{P1,P2,P3,P4}. Unlike standard Mpemba comparisons that involve different initial conditions under a fixed generator, the PME compares distinct dynamical protocols starting from the \emph{same} initial state and ending at the same stationary state. A typical two-step Pontus protocol initially drives the system under a modified generator to an auxiliary state and then lets it relax under the target generator; if the total time is shorter than that of the direct protocol, the PME is observed. This protocol-based mechanism was theoretically analyzed in a range of classical and quantum settings and has been argued to reveal the role of non-normal spectral geometry and intermediate-state engineering in accelerating relaxation~\cite{r28,P1,P2,P3,P4}.

Concurrently, non-Hermitian physics has uncovered a number of surprising phenomena in open and effectively non-Hermitian lattice systems \cite{Ashida2020}, notably the \emph{non-Hermitian skin effect} (NHSE) \cite{NH1,NH2,NH3,NH4,NH5,NH6,NH7,NH8,NH9,NH10,NH11,NH12,NH13,NH14,NH15,NH15b,NH16,NH17,NH18,NH19,NH20,NH21,NH22,NH23} whereby eigenmodes pile up at system boundaries under non-reciprocal couplings, invalidating conventional bulk--boundary correspondence and producing extreme boundary sensitivity.
 For open quantum systems, non-reciprocal dissipation can similarly induce a Liouvillian skin effect (LSE) \cite{NH7,LS1,LS2,LS3,LS4,LS5,LS6}: left and right eigenmodes of the Liouvillian superoperator become localized at opposite edges, producing strong non-orthogonality~\cite{NH7,LS1}. The LSE and similar non-normal spectral features have been connected to long transients and sensitivity to boundary conditions, slow relaxation and anomalous transport in a number of recent works~\cite{NH7,LS1,LS2,LS3}.

This paper aims to bridge these two emergent areas. Building on prior quantum-Mpemba literature and recent Pontus studies, we show that the Liouvillian skin effect can enable a Pontus--Mpemba acceleration by reducing the projection onto the slowest Liouvillian component. We analyze a minimal number-conserving open quantum chain with coherent nearest-neighbour hopping $J$, asymmetric incoherent hoppings $J_{\rm L}, J_{\rm R}$, and a tunable coherent boundary coupling $\epsilon$. For $\epsilon = 0$ and $J_{\rm R} \neq J_{\rm L}$, the Liouvillian exhibits the skin effect, with right eigenmodes exponentially localized at one edge and left eigenmodes at the opposite edge. A two-step Pontus protocol that first coherently transfers the excitation from one edge to the opposite one prepares an auxiliary state in which the projection on the slow boundary-localized Liouvillian mode is strongly reduced, though not entirely eliminated. This partial suppression leads to an accelerated transient relaxation, while the asymptotic decay rate of the system remains unchanged. We provide analytic reasoning in the single-excitation sector and numerical results that illustrate this mechanism. The results presented here offer a new perspective on controlling relaxation dynamics in open quantum systems by exploiting non-Hermitian effects such as the Liouvillian skin effect. Beyond their fundamental significance for understanding anomalous transient behavior, these findings may inform the design of quantum devices and protocols where fast initialization, state preparation, or dissipation engineering is desired. Moreover, the combination of geometric mode localization and targeted coherent manipulation provides a versatile framework that could inspire further investigations in many-body systems, quantum thermodynamics, and non-equilibrium statistical mechanics, opening avenues for both theoretical exploration and experimental realization.

\section{Model and Liouvillian dynamics}
A schematic of the dissipative model considered in this work is illustrated in Fig.1(a) \cite{LS1,LS3}. The system consists of a one-dimensional tight-binding chain of $L$ sites with the coherent dynamics described by the Hamiltonian 
\begin{equation}
H = -J\sum_{n=1}^{L-1} (a_{n+1}^\dagger a_n + a_n^\dagger a_{n+1}) - \epsilon (a_1^\dagger a_L + a_L^\dagger a_1),
\label{eq:H}
\end{equation}
where $J$ is the nearest-neighbour hopping rate, $\epsilon$ is a tunable coherent coupling between the end sites, and $a_n^{\dag}$, $a_n$ are the creation and annihilation operators of a boson at
site $n$, which satisfy the usual bosonic commutation relations. The dissipative dynamics is obtained by including asymmetric left/right incoherent hopping at rates $J_L$ and $J_R$, which are implemented by Lindblad jump operators \cite{LS1,H1,H2,H3}
\begin{equation}
L_n^{(R)}=\sqrt{J_{\rm R}}\,a_{n+1}^\dagger a_n,\qquad L_n^{(L)}=\sqrt{J_{\rm L}}\,a_{n}^\dagger a_{n+1},\qquad n=1,\dots,L-1.
\label{eq:jumpops}
\end{equation}

This model can be physically implemented with ultracold atoms in an optical lattice by
laser-assisted hopping with spontaneous emission \cite{LS1}, photonic quantum walks in
synthetic lattices \cite{IMP1,IMP2,IMP3} and models of asymmetric simple
exclusion processes in bosonic lattices \cite{IMP4}. 
Within the
Born-Markov approximation, the time evolution of
the density matrix $\rho$ of the system is described by the Lindblad master equation (see e.g. \cite{LS1})
\begin{equation}
\dot{\rho}=\mathcal{L}\rho = -i[H,\rho] + \sum_{n=1}^{L-1}\Big( \mathcal{D}[L_n^{(R)}]\rho + \mathcal{D}[L_n^{(L)}]\rho\Big),
\label{eq:L}
\end{equation}
\begin{figure}[t]
\centering
\includegraphics[width=0.6\textwidth]{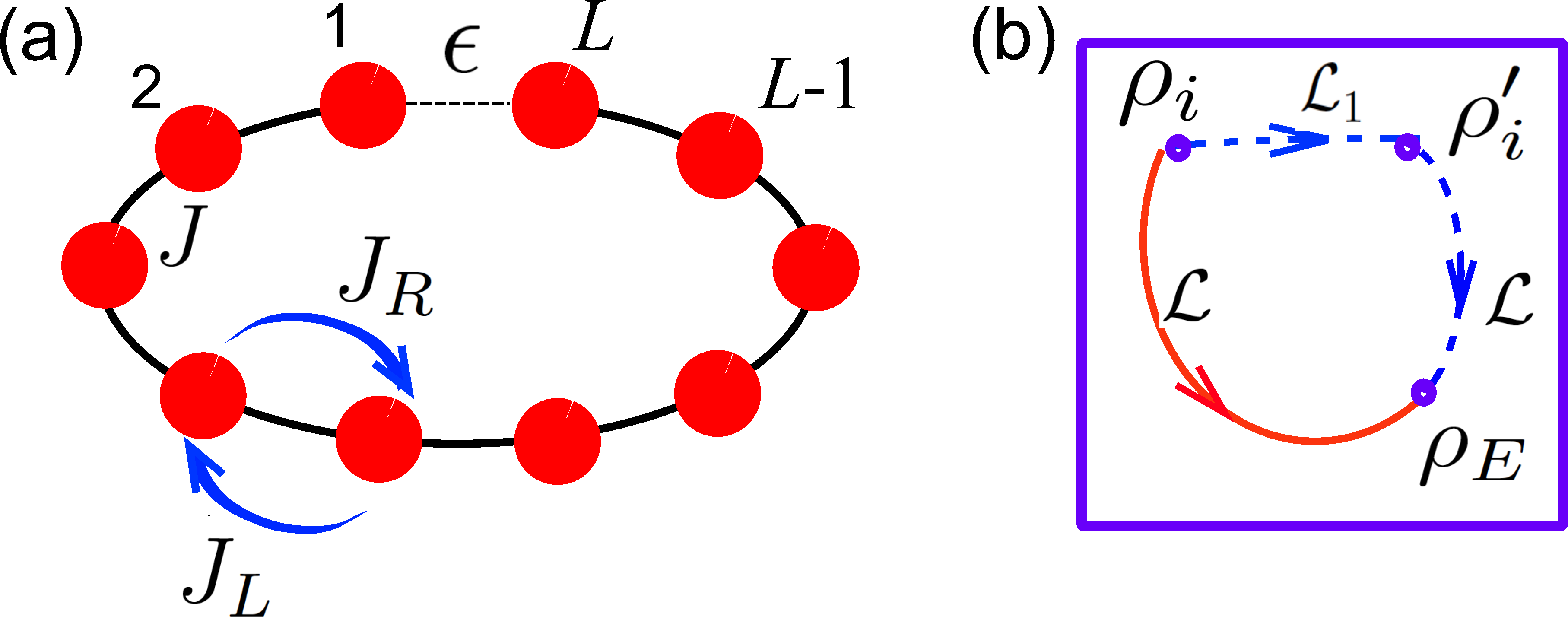}
\caption{(a) Schematic of the tight-binding chain made of $L$ sites with coherent ($J$) and asymmetric incoherent($J_R, J_L$)  hoppings between adjacent sites. The end sites $n=1$ and $n=L$ of the chain are connected by a coherent hopping at tunable rate $\epsilon$. (b) Illustration of the quantum Pontus-Mpemba effect in phase space. The initial state $\rho_i$  can relax toward the same stationary state $\rho_E$ via two protocols.  In the single-step (direct) protocol the system evolves under the Liouvillian $\mathcal{L}$.  In the two-step (Pontus) protocol, for $0\le t\le \tau$ the system is evolved under a preparatory Liouvillian $\mathcal{L}_1$ that drives the  system toward the auxiliary state $\rho_i^{\prime} $. For $t>\tau$ the system then evolves under the same Liouvillian $\mathcal{L}$ as in the single-step protocol.  Both protocols approach the same stationary state $\rho_E$. The Pontus-Mpemba effect arises whenever the time to reach the stationary state is shorter in the two-step protocol.}
\label{fig1}
\end{figure}
with $\mathcal{D}[O]\rho = O\rho O^\dagger - \frac{1}{2}\{O^\dagger O,\rho\}$. 
 The evolution generated by the Liouvillian $\mathcal{L}$ preserves the
total number of bosons $N=\sum_n a^{\dag}_n a_n$, which constitutes a strong symmetry of the system, {\color{black}{i.e. $N$ commutes \emph{both} with the Hamiltonian and with all Lindblad jump operators.} } If we restrict its action to
a Hilbert space sector with a fixed number of bosons, then
its stationary (equilibrium) state $\rho_E$ , corresponding to $\mathcal{L} \rho_E=0$, is unique \cite{LS3}. In the sector of $N$ excitations, the dimension of Hilbert space is $D=(N+L-1)! /[ N! (L-1)!]$, the density operator can be represented by a vector of size $D^2$ and the Liouvillian $\mathcal{L}$  by a square matrix of size $D^2 \times D^2$. 
For a given initial state $\rho_i$, the relaxation dynamics toward the steady-state is governed by the eigenvalues and corresponding right and left eigenvectors of the Liouvillian $\mathcal{L}$.
Let $\{\lambda_\alpha\}$ denote the eigenvalues of $\mathcal{L}$, with associated right and left eigenvectors $R_\alpha$ and $L_\alpha$ satisfying biorthogonality $ \langle L_\alpha|R_\beta \rangle={\rm Tr} ( L_\alpha^{\dag} R_\beta)= \delta_{\alpha\beta}$, with $ \alpha, \beta=1,2,3,..., D^2$. The eigenvalues are ordered such that $\lambda_1=0> {\rm Re}(\lambda_2) \geq {\rm Re}(\lambda_3) \geq ... \geq {\rm Re} (\lambda_{D^2}).$ In the absence of spectral singularities (exceptional points), the initial state can be expanded on the right eigenvector eigenbasis as
\begin{equation}
\rho_i = \rho_E + \sum_{\alpha>1} c_\alpha R_\alpha,\qquad c_\alpha = \langle L_\alpha|\rho_i \rangle= {\rm Tr} \left( L_{\alpha}^{\dag} \rho_i \right).
\label{eq:expansion}
\end{equation}
The time evolution then reads
\begin{equation}
\rho(t) = \rho_E + \sum_{\alpha>1} c_\alpha e^{\lambda_\alpha t} R_\alpha.
\label{eq:timeexp}
\end{equation}
The relaxation of the system toward the equilibrium state $\rho_E$ can be quantified using standard measures of distance between density matrices. One common choice is the {trace distance}, defined as
\begin{equation}
D_{\rm tr}(\rho, \rho_E) = \frac{1}{2} \, \mathrm{Tr}\, |\rho - \rho_E|,
\end{equation}
where $|A| = \sqrt{A^\dagger A}$. Another frequently used measure is the {Hilbert--Schmidt distance}, given by
\begin{equation}
D_{\rm HS}(\rho, \rho_E) = \sqrt{\mathrm{Tr}\, \big(\rho - \rho_E\big)^2}.
\end{equation}
Both distances are positive and vanish if and only if $\rho = \rho_E$, providing a natural way to monitor how close the system is to equilibrium during its dissipative evolution. Equation~(5) indicates that, in general, an initial state with a smaller overlap (i.e., a smaller modulus of $c_1$) with the slowest-decaying eigenvector $R_2$ will relax to equilibrium more rapidly than one with a larger overlap. An interesting property of the model is that, for asymmetric incoherent hopping ($J_L \neq J_R$), it exhibits the Liouvillian skin effect, that is, a pronounced sensitivity of the spectrum and of the left and right eigenvectors to the boundary coupling $\epsilon$~\cite{LS1,LS3}. Specifically, for $\epsilon = 0$, the right and left eigenvectors of the Liouvillian $\mathcal{L}$ are exponentially localized near opposite edges of the lattice.  The Liouvillian skin effect  has been associated  to slowing down of relaxation processes without gap closing due to non-normality of the Liouvillian \cite{LS1}.
As we will show in the next section, the Liouvillian skin effect arising from asymmetric incoherent hopping can give rise to the emergence of the Pontus-Mpemba effect as well, i.e. to accelerated relaxation to stationarity in a two-step protocol.\\
In this work, we will focus on the single-particle sector $N=1$, which captures the essential features of the relaxation dynamics. 
Restricting to the single-excitation manifold spanned by the basis $\{|n\rangle=a_n^\dagger|0\rangle\}$ with $n=1,2,...,L$, the density matrix elements are $\rho_{mn}=\langle m|\rho|n\rangle$. Using the Hamiltonian~(\ref{eq:H}) and the jump operators~(\ref{eq:jumpops}), one readily obtains the following evolution equations for the density matrix elements
\begin{eqnarray}
\dot{\rho}_{n,m} & = &  i \sum_{k=1}^L \left( \rho_{n,k}H_{k,m}-H_{n,k} \rho_{k,m} \right) \nonumber \\
& + & J_R \delta_{n,m} (1- \delta_{n,1}) \rho_{n-1,n-1}+ J_L \delta_{n,m} (1-\delta_{n,L}) \rho_{n+1,n+1} \\
& - & (J_R+J_L) \rho_{n,m}+ \frac{1}{2}J_R (\delta_{n,L}+\delta_{m,L}) \rho_{n,m} + \frac{1}{2}J_L (\delta_{n,1}+\delta_{m,1}) \rho_{n,m} \nonumber
\label{eq:rho_mn}
\end{eqnarray}
where 
\begin{equation}
H_{n,m}= \langle n|  H | m \rangle = \left(
\begin{array}{cccccccc}
0 & -J & 0 & 0 & ... & 0 & 0 & -\epsilon \\
-J & 0 & -J & 0 & ...& 0 & 0 & 0 \\
... & ... & ... & ... & ... & ... & ... & ... \\
0 & 0 & 0 & 0 & ... & -J & 0 & -J \\
-\epsilon & 0 & 0 & 0 & ... & 0 & -J & 0
\end{array}
\right).
\end{equation}
\begin{figure}[t]
\centering
\includegraphics[width=0.95\textwidth]{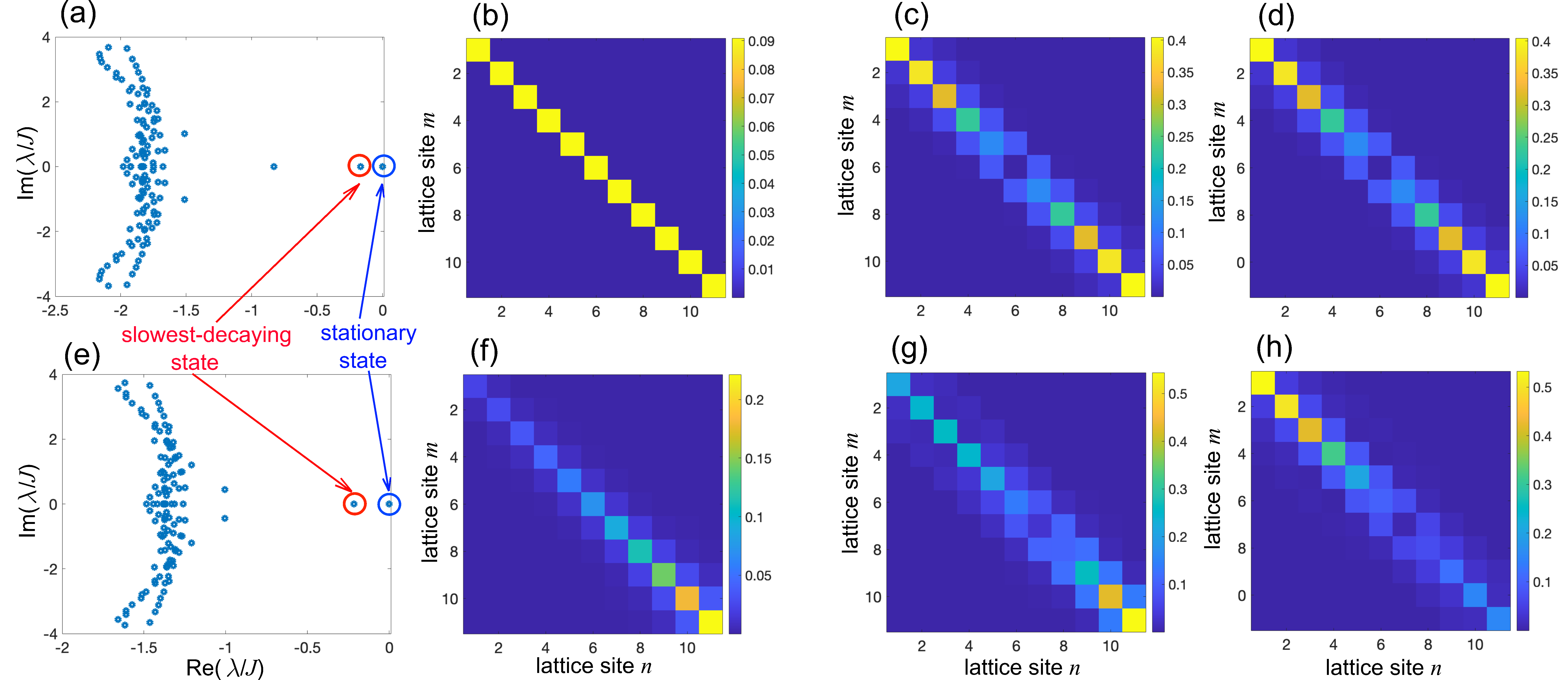}
\caption{(a) Spectrum of the Liouvillian $\mathcal{L}$ in the single-excitation subspace for parameter values $L=11$, $\epsilon=0$, $J_R/J=1$ and $J_L/J_R=1$ (symmetric hopping). The eigenvalues $\lambda_1=0$ and $\lambda_2$, corresponding to the stationary (equilibrium) state $\rho_E$ and to the slowest-decaying state, are highlighted with blue and red circles, respectively. (b) Behavior of the stationary state [modulus of $(\rho_E)_{n,m}$]. (c,d) Behavior of the right eigenvector $R_2$ [modulus of $(R_2)_{n,m}$, panel (c)] and left eigenvector  $L_2$ [modulus of $(L_2)_{n,m}$, panel (d)], corresponding to the eigenvalue $\lambda_2$.   (e-h) Same as (a-d), but for $J_L/J_R=0.5$ (asymmetric hopping).}
\label{fig2}
\end{figure}
Equation~(\ref{eq:rho_mn}) explicitly displays how populations and coherences couple through coherent currents (proportional to $J,\epsilon$) and incoherent asymmetric transfers ($J_{\rm R},J_{\rm L}$).  A representative example of numerically computed spectra of the Liouvillian $\mathcal{L}$ for $\epsilon = 0$ and for both symmetric ($J_L = J_R$) and asymmetric ($J_R > J_L$) incoherent hopping are shown in Fig.~2, together with the profiles of the stationary state $\rho_E$ and of right and left eigenvectors, $R_2$ and $L_2$, corresponding to the slowest-decaying mode ($\alpha = 2$). 
In the symmetric case $J_L=J_R$, the stationary state is independent of $J$, diagonal in the $\{ |n \rangle \}$ basis and given by $\rho_E=(1/L){\rm diag} (1,1,..., 1)$. Conversely, in the asymmetric case ($J_L \neq J_R$) the equilibrium state displays non-vanishing coherences ($(\rho_E)_{n,m} \neq 0$ for $n \neq m$). Furthermore, as one can see from Fig.2 asymmetry in the incoherent hopping rates leads to an exponential localization of  both the stationary state and of right/left eigenvectors $R_2, L_2$ near lattice edges, with opposite edges for left and right eigenvectors, which is a clear signature of the Liouvillian skin effect. 
{\color{black} A special limiting case is obtained by letting $\epsilon=J=0$, where the quantum master equation (8) reduces to a classical continuous--time birth--death
stochastic process on a chain of $L$ sites with reflecting boundaries. In this case, all off--diagonal
density--matrix elements $\rho_{n,m}(t)$ with $n\neq m$ decay to zero, and for any initial condition with vanishing coherences, the dynamics is restricted to the diagonal diagonal elements
\(P_n(t) = \rho_{n,n}(t)\) (populations), which satisfy 
satisfy a closed classical birth-depth evolution equation with rates $J_L$ and $J_R$. This case is discussed in the Appendix A. The stationary probability distribution is given by
\[
\rho_E= \mathcal{N} {\rm diag} \left (1,r,r^2,...,r^{L-1}  \right)
\]
where 
\[
r \equiv \frac{J_R}{J_L} \nonumber
\]
is the skin parameter and $\mathcal{N}=1 / \sum_{n=0}^{L-1} r^n$ is the normalization factor.  We are specifically interested to study the relaxation dynamics for the following two initial conditions:
\begin{equation}
\rho_i = |1\rangle\langle 1|,
\end{equation}
corresponding to excitation at the end lattice site $n=1$, or
\begin{equation}
\rho_i^{\prime} = |L\rangle\langle L|,
\end{equation}
corresponding to excitation at the end lattice site $n=L$. As shown in the Appendix A, for large system size $L$ and for $\alpha \ll L$, the ratio of the spectral amplitudes entering in Eq.(5) satisfy the simple condition
\begin{equation}
\left|  \frac{c_{\alpha}}{c_{\alpha}^{\prime}} \right| = r^{L-1}
\nonumber
\end{equation}
clearly indicating that, for $ r \neq 1$, one of the two initial states relaxes toward the equilibrium distribution much slower than the other one, since it excites much more than the other condition the slowing-decaying states. For example, assuming $J_R>J_L$, i.e. $r>1$, the initial state $\rho_i$ relaxes to equilibrium much slower than the initial state $\rho^{'}_i$. Physically, this behavior has a very simple explanation: for $r>1$, the excitation at equilibrium is mostly localized at the right edge $n=L$ of the chain, and thus $\rho_i^{\prime}$ is closer to and relaxes faster toward the equilibrium distribution.\\ For $J \neq 0$, i.e. when hopping in the lattice is due to both incoherent (classical) and coherent (quantum) processes, analytical results are not available, however the same physical picture for the asymmetry in the relaxation dynamics is retained. 
}

\section{Pontus--Mpemba effect induced by the skin effect}
To illustrate how the Liouvillian skin effect can be harnessed to observe the Pontus-Mpemba effect, let us first consider a single-step protocol, where the systems relaxes toward the stationary state $\rho_E$ for $\epsilon=0$ and fixed values of $J$, $J_R$ and $J_L=J_R/r$. Let us prepare the system in either one of the two pure states
$\rho_i = |1\rangle\langle 1|$,
corresponding to excitation at the end lattice site $n=1$, or
$\rho_i^{\prime} = |L\rangle\langle L|$,
\begin{figure}[t]
\centering
\includegraphics[width=0.7\textwidth]{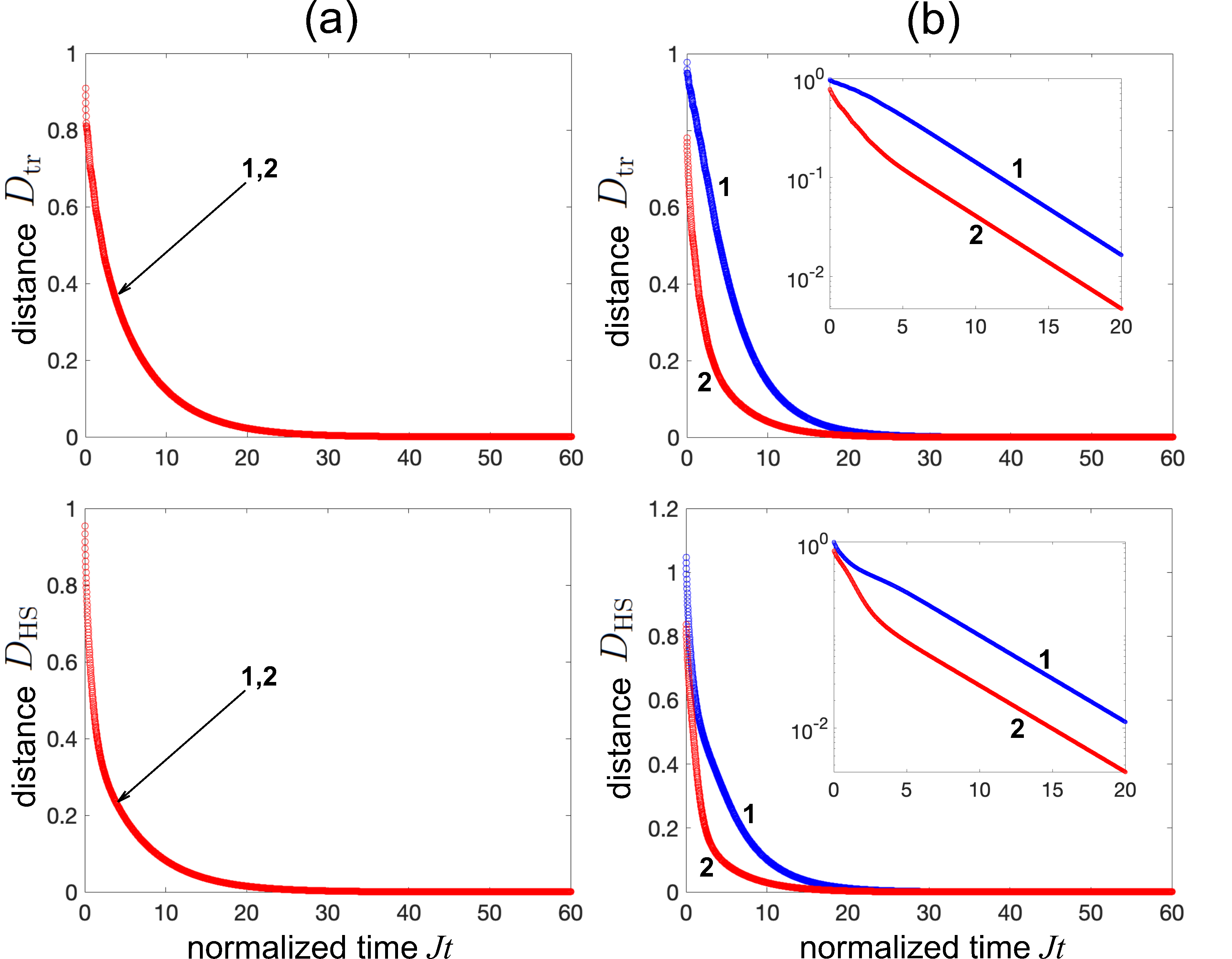}
\caption{Relaxation dynamics (temporal evolution of the trace distance $D_{tr}$, upper plots, and Hilbert-Schmidt distance $D_{HS}$, lowers plots) under the Liouvillian $\mathcal{L}$ for parameter values $\epsilon=0$, $J_R/J=1$, $L=11$ and for (a) $J_L/J_R=1$ (symmetric incoherent hopping) and (b) $J_L/J_R=0.5$ (asymmetric incoherent hopping). Curves 1 and 2 refer to different initial states (curve 1: $\rho_i=|1 \rangle \langle 1|$, curve 2: $\rho_i^{\prime}=|L \rangle \langle L|$). In the symmetric hopping case owing to mirror symmetry curves 1 and 2 are overlapped yielding the same relaxation dynamics. The insets in panels (b) show the relaxation dynamics on a vertical log scale.}
\label{fig3}
\end{figure}
\begin{figure}[t]
\centering
\includegraphics[width=0.7\textwidth]{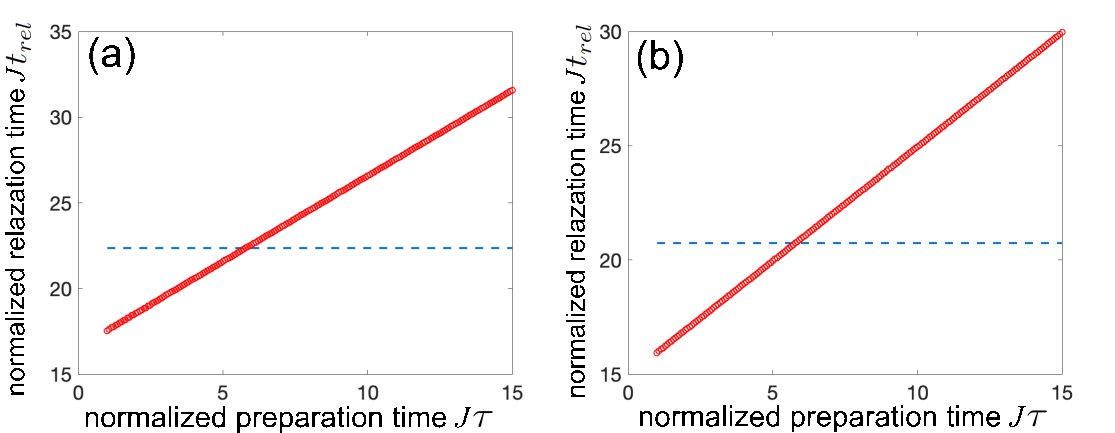}
\caption{Numerically-computed relaxation time $t_{rel}$ versus preparation time $\tau$ in the two-step (Pontus) protocol (red circles). The horizontal dashed blue curve is the relaxation time, independent of $\tau$, in the one-step (direct) protocol. Parameter values are as in Fig.3(b). The relaxation time $t_{rel}$ is defined as the time instant such that $D(\rho(t_{tr}), \rho_E)=0.01$. In panel (a), the distance $D$ is the trace distance, whereas in panel (b) the distance $D$ is the Hilbert-Schmidt distance.}
\label{fig4}
\end{figure}
corresponding to excitation at the end lattice site $n=L$. Clearly, when $J_L = J_R$, the symmetry of the system ensures that the relaxation toward equilibrium, as measured by either the trace distance or the Hilbert--Schmidt distance, is identical for the two initial states; see Fig.~3(a) for an illustrative example. However, when asymmetry is introduced by setting, for instance, $J_R > J_L$, i.e. $r>1$, as discussed at the end of the previous section the 
state $\rho_i$ is farther away than $\rho_i^{\prime}$ from the equilibrium state $\rho_E$, and the overlap of the initial state $\rho_i$ with the slowest-decaying eigenvector---that is, the spectral coefficient $c_2$---becomes noticeably larger than that of the initial state $\rho_i^{\prime}$ due to boundary localization (the Liouvillian skin effect). This implies that, although the last asymptotic decay rate is the same and given by $|{\rm  Re}(\lambda_2)|$, the relaxation toward equilibrium occurs more rapidly for the initial state $\rho_i^{\prime}$ than for $\rho_i$; see Fig.~3(b) for an illustrative example. In particular, as shown in the insets of Fig.3(b) the relaxation in the early stage is markedly faster for state $\rho_i^{\prime}$ than for $\rho_i$. \textcolor{black}{This effect clearly depends on the system size $L$, and owing to Eq.(12) it becomes more and more pronounced as $L$ is increased, with an extremely slow relaxation of initial state $\rho_i$, i.e. almost frozen dynamics for the distance $D(t)$, in the thermodynamic limit $L \rightarrow \infty$. The physical reason is simple: for $r = J_R/J_L > 1$, the stationary excitation profile is strongly localized at the right edge $n = L$ of the chain. Consequently, $\rho_i'$ starts much closer to the stationary distribution and relaxes more rapidly. In contrast, in the large-$L$ limit, relaxation from $\rho_i$ requires the excitation to propagate across the entire chain; because transport occurs at a finite speed, the time needed to reach the opposite boundary grows with $L$, leading to an increasingly slow relaxation.
 However, we remark that the effect does not require to assume the thermodynamic limit and, as shown in Fig.3, it is well visible even for a relatively small value of $L$.}

 The above results can be exploited to realize a quantum Pontus-Mpemba effect. We consider two protocols starting from the same initial pure state $\rho_i$ given by Eq.(10); see Fig.1(b) for a schematic. In the single-step (direct) protocol the system evolves under $\mathcal{L}$ of Eq.~(\ref{eq:L}) with $\epsilon=0$, until the stationary state $\rho_E$ is reached, as already illustrated by the blue curves in Fig.3(b).  In the two-step (Pontus) protocol, for $0\le t\le \tau$ the system is evolved under a preparatory Liouvillian $\mathcal{L}_1$ obtained by setting $J_{\rm R}=J_{\rm L}=J=0$ and $\epsilon=\epsilon_1$ in Eq.~(\ref{eq:L}), so that the evolution is purely coherent and implements a swap (Rabi flopping) between sites $1$ and $L$ when $\tau=\pi/(2\epsilon_1)$. The auxiliary state $\rho_i^{\prime} =|L\rangle\langle L|$ is thus prepared in a time $\tau$. For $t>\tau$ the system evolves under the same Liouvillian $\mathcal{L}$ as in the single-step protocol (with $\epsilon=0$).  Both protocols approach the same stationary state $\rho_E$.
Quantifying the relaxation by either trace distance or Hilbert--Schmidt norm, we then compare the times $t_{rel}$ required by the two protocols to reach a given threshold distance. As an example, Fig.4 shows the numerically-computed relaxation time $t_{rel}$ in the two protocols, where $t_{rel}$ is defined such that either the trace distance $D_{tr}$ or Hilbert-Schmidt distance $D_{HS}$ reaches the threshold value 0.01. Clearly, the relaxation time is independent of $\tau$ in the one-step (direct) protocol, while it turns out to increase with $\tau$ in the two-step (Pontus) protocol. As Fig.4. clearly shows, provided that the preparation time $\tau= \pi /( 2 \epsilon_1)$ is shorter than  $\sim 5/J$, the two-step (Pontus) protocol reduces the time for equilibration as compared to the single-step (direct) protocol; see Fig.5 for an illustrative example that compares the relaxation dynamics for the two protocols.  The accelerated effect vanishes when $J_L=J_R$ (Fig.6), clearly indicating that the Pontus-Mpemba effect is enabled by the Liouvillian skin effect. 

\begin{figure}[t]
\centering
\includegraphics[width=0.7\textwidth]{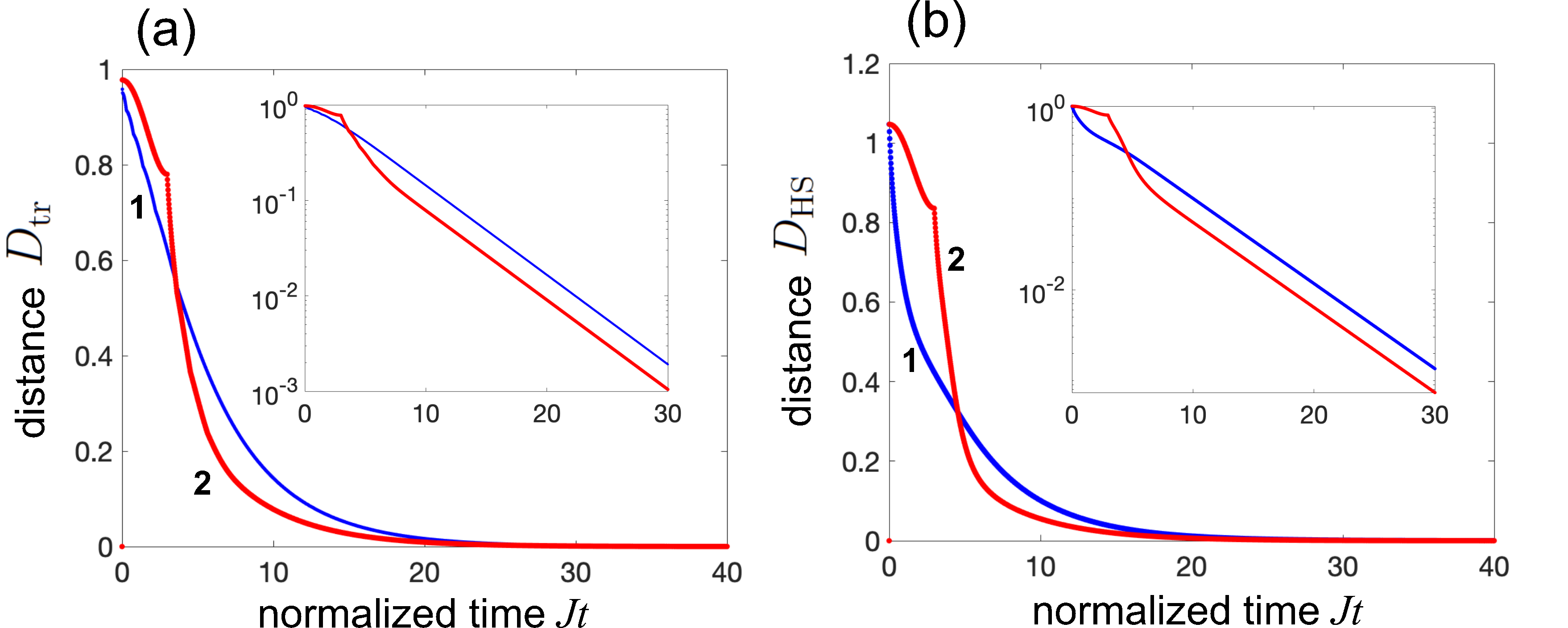}
\caption{Illustrative relaxation dynamics showing the emergence of the Pontus-Mpemba effect. Panels (a) and (b) display the numerically-computed temporal evolution of the trace distance $D_{tr}(\rho(t), \rho_E)$ [panel (a)] and the Hilbert-Schmidt distance $D_{HS}(\rho(t), \rho_E)$ [panel (b)] for the one-step (direct) protocol (curve 1) and the two-step (Pontus) protocol (curve 2). Parameter values of the Liouvillian $\mathcal{L}$ are $\epsilon=0$, $J_R=J$, $J_L/J_R=0.5$, $L=11$. In the first-step of the Pontus protocol, the Liouvillian is $\mathcal{L}_1$ with $J=J_L=J_R=0$ and $\epsilon=\epsilon_1= \pi/(2 \tau)$ with $\tau=3$ (in units of $1/J$ of first-step protocol). The insets show the relaxation dynamics on a vertical log scale.}
\label{fig5}
\end{figure}

\begin{figure}[t]
\centering
\includegraphics[width=0.7\textwidth]{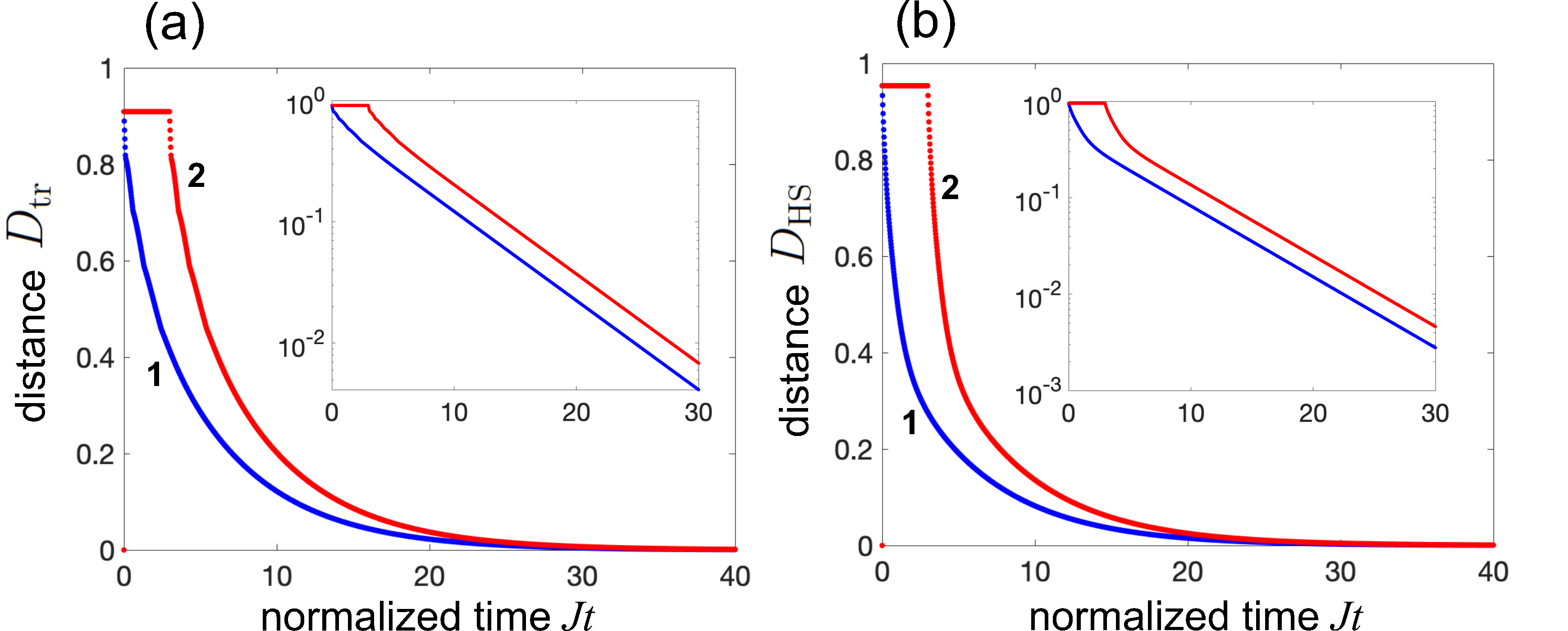}
\caption{ Same as Fig.5, but for $J_L/J_R=1$.}
\label{fig5}
\end{figure}

\section{Discussion and conclusions}

The Pontus--Mpemba acceleration described here arises as a geometric consequence of Liouvillian non-normality. The Liouvillian skin effect induces boundary-localized left and right eigenmodes, and a suitable coherent preconditioning can shift the initial state toward the boundary where the slow left eigenmode has minimal support. While the  asymptotic decay rate remains unchanged, the reduced overlap with the slow mode leads to a pronounced transient speedup. This mechanism highlights a direct link between non-normal spectral geometry and protocol-dependent relaxation dynamics, connecting recent advances in Mpemba physics with the growing field of non-Hermitian and Liouvillian skin effects. {\color{black}{From an experimental perspective, the ingredients of the model presenetd in this work are well within reach of current platforms. In ultracold atoms, asymmetric incoherent hopping with tunable rates $J_L$ and $J_R$ can be engineered using laser-assisted tunnelling combined with directional optical pumping, while coherent boundary coupling $\epsilon$ may be realized via Raman-assisted long-range hopping or by coupling the edge sites through an ancillary state or auxiliary lattice. Synthetic photonic lattices provide another promising route, where dissipative coupling and controlled asymmetry can be introduced using optical loss engineering, electro-optic modulation, or non-reciprocal waveguide couplers. In both settings, the preparation of the initial single-excitation state---either localized at a boundary or transferred coherently from the opposite edge---is standard and requires no fine-tuning beyond typical state-preparation protocols. Finally, it should be mentioned that  the Pontus--Mpemba acceleration described here is expected to be robust against additional noise and dephasing. In fact, as discussed in the previous sections the effect is observable even in the fully classical limit of the quantum master equation. In this regime the dynamics reduces to a classical biased birth--death process, for which the Mpemba-like behaviour arises solely from the asymmetric dissipative rates. 
} } Possible extensions of this work include exploring higher excitation sectors, disordered or interacting chains, and optimizing the preparatory stage to minimize the overlap with the slowest Liouvillian component.  {\color{black}{In the many-particle regime, where multiple excitations and interactions come into play, the qualitative picture outlined here is expected to extend in spirit but with richer phenomenology. Interactions generally modify both the structure of the Liouvillian spectrum and the spatial profiles of its slow modes, potentially giving rise to collective or density-dependent skin effects. As a consequence, the geometry of overlaps that governs the Pontus--Mpemba acceleration may acquire an intrinsically many-body character, with slow Liouvillian modes becoming dressed by interaction-induced correlations. Although a full analytical treatment is challenging, we expect that suitably chosen preparatory protocols could still suppress the overlap with these slow collective modes, enabling Mpemba-like acceleration in interacting systems. Investigating these questions, and identifying regimes where interactions enhance or hinder the effect, represents an interesting direction for future work.}}

Overall, our results demonstrate that the Liouvillian skin effect can serve as a resource for engineering accelerated relaxation protocols in open quantum systems, offering new insights into the interplay between non-Hermitian geometry and nonequilibrium dynamics.

\section*{Acknowledgments}
The author acknowledges the Agencia Estatal de Investigacion (MDM-2017-0711).

\appendix
{\color{black}{
\section*{Appendix: Classical birth--death dynamics from the quantum model}

In this appendix we show that, when the quantum model is restricted to the
single--excitation sector and all coherent couplings vanish ($\epsilon = J = 0$),
the master equation describing purely dissipative dynamics reduces to a classical continuous--time birth--death
process on a chain of $L$ sites with reflecting boundaries. In this limiting case, 
one can derive analytically the stationary state, the spectrum of the generator, the right and
the left eigenvectors, clearly identifying the key role of the Liouvillian skin effect in the relaxation dynamics.

\subsection*{A. Reduction to a classical master equation}

In the single--excitation basis $\{ |1\rangle, |2\rangle, \ldots, |N\rangle \}$,
and for $\epsilon = J = 0$, the Hamiltonian vanishes and all off--diagonal
density--matrix elements $\rho_{n,m}(t)$ with $n\neq m$ decay to zero.
The diagonal elements
\[
P_n(t) = \rho_{n,n}(t)
\]
satisfy a closed classical birth--death stochastic process, namely.
\begin{eqnarray}
\frac{dP_1}{dt} & = & J_L P_2 - J_R P_1 \\
\frac{dP_n}{dt}
& = &  J_L P_{n+1} + J_R P_{n-1}
- (J_L + J_R) P_n  \; \; (n=2,3,..,L-1)) \\
\frac{dP_L}{dt} & = &  J_R P_{L-1} - J_L P_L .
\end{eqnarray}
In vector form,
\[
\frac{d\mathbf{P}}{dt} = M\,\mathbf{P},
\]
where $M$ is the $N\times N$ birth--death Markov generator.

\subsection*{B. Stationary state}

The stationary (equilibrium) distribution $\mathbf{P}^{(E)}$ satisfies \(
M \mathbf{P}^{(E)} = 0  \) and can be readily computed by the detailed balance condition, yielding
\[
P^{(E)}_n =  \mathcal{N} \, r^{\,n-1},
\]
where $r=J_R/J_L$ and $\mathcal{N}=1 / \sum_{n=0}^{L-1} r^n$ is the normalization factor.
For $r=1$ the distribution is uniform, while for $r \neq 1$ it is squeezed toward one of the two lattice edges, a characteristic signature of the Liouvillian skin effect.

\subsection*{C. Eigenvalues and Right/Left eigenvectors}
The eigenvalues $\lambda_{\alpha}$ and corresponding right $\mathbf{R}^{(\alpha)}$ and left $\mathbf{L}^{(\alpha)}$ eigenvectors of the Markov transition matrix $M$ can be calculated analytically as follows.
Define the diagonal matrix
\[
D = \mathrm{diag}(1, r, r^2, \ldots, r^{N-1}).
\]
The matrix $U$ obtained via the similarity transformation
\[
U = D^{-1/2} M D^{1/2}
\]
is real and symmetric (Hermitian), namely one has
\begin{equation}
U=\left(
\begin{array}{cccccccc}
-J_R & J & 0 & 0 & ... & 0 & 0 & 0 \\
J & -(J_R+J_L) & J & 0 & ... & 0 & 0 & 0 \\
... & ... & ... & ... & ... & ... & ... & ... \\
 0 & 0 & 0 & 0 & ... & J & -(J_R+J_L) & J \\
 0 & 0 & 0 & 0 & ... & 0 & J & -J_L \\
\end{array}
\right)
\end{equation}
where we have set $J \equiv \sqrt{J_RJ_L}$. Hence $M$ is diagonalizable with real eigenvalues. Let us indicate by  $\mathbf{V}^{(\alpha)}$ and $\lambda_{\alpha}$ the eigenvectors and corresponding eigenvalues of the Hermitian lattice matrix $U$. Clearly, $M$ has the same eigenvalues than $U$, and the right/left  eigenvectors of $M$ are obtained from $\mathbf{V}^{(\alpha)}$ via the transformations
$\mathbf{R}^{(\alpha)}=D^{1/2}\mathbf{V}^{(\alpha)}$ and $\mathbf{L}^{(\alpha)}=D^{-1/2}\mathbf{V}^{(\alpha)}$, i.e.
\begin{equation}
R_n^{(\alpha)}=r^{n-1} V_n^{(\alpha)} \; ,\;\;  L_n^{(\alpha)}=\left(\frac{1}{r}\right)^{n-1}  V_n^{(\alpha)}.
\end{equation}
Since $V_n^{(\alpha)}$ are oscillating functions of $n$ on the lattice, Eq.(17) clearly shows the squeezing of the right and left eigenvectors of $M$ toward the two opposite lattice edges, which is the clear signature of the skin effect. The eigenvectors $\mathbf{V}^{(\alpha)}$ of $U$ and corresponding eigenvalues $\lambda_{\alpha}$ can be analytically calculated from the Bethe Ansatz
\begin{equation}
V_n^{(\alpha)}= \left\{
\begin{array}{ll}
S_1 & n=1 \\
A e^{ik(n-2)}+Be^{-ik(n-2)} & 2 \leq n \leq L-1 \\
S_L & n=L
\end{array}
\right.
\end{equation}
where 
\begin{equation}
\lambda=-J_L-J_R+2 J \cos k
\end{equation}
and $k$ is a (generally-complex) wave number which is determined from a solvability condition obtained by imposing the Ansatz (18) to satisfy the lattice equation at $n=1,2$ and $n=L-1,L$ sites. The solvability condition reads explicitly
\[
\left( -J_R-J_L+2J \cos k  \right) e^{2ikL}=1
\]
which is solved by letting either $e^{2ikL}=1$ or $-J_R-J_L+2J \cos k=0$. The latter condition yields a complex $k$ value and corresponds to the zero eigenvalue $\lambda_1=0$ of the stationary state of the Markov generator matrix $M$. The other $(L-1)$ eigenvalues, obtained by letting $\exp(2ikL)=1$, i.e. $k= (\alpha-1) \pi /L$, correspond to the decaying modes and read
\begin{equation}
\lambda_{\alpha}=-J_L-J_R+2 \sqrt{J_L J_R} \cos \left(  \frac{\pi (\alpha-1)}{L} \right)
\end{equation}
with $\alpha=2,3,...,L$. From the boundary conditions analysis it readily follows that the ratio $S_L/S_1$ of the eigenmode amplitudes $S_1$ and $S_L$ at the two lattice edges is given by
\[
\left| \frac{S_1}{S_L} \right|= \left|  \frac{A+B}{A e^{3ik}+Be^{-3ik} } \right|.
\]
From the above relation, it follows that, in the large $L$ limit and for $\alpha \ll L$, such that $k=  (\alpha-1) \pi/ L \ll 1$, one has $|S_1/S_L| \simeq 1$.

\subsection*{D. General relaxation dynamics}

Let us assume that the left and right eigenvectors of $M$ are normalized such as to satisfy the biorthogonal condition $\langle \mathbf{L}^{(\alpha)} | \mathbf{R}^{(\beta)} \rangle= \delta_{\alpha,\beta}$, and let us
expand the initial probability vector as a sum of the right eigenvectors of the Markov generator $M$, i.e.
\[
\mathbf{P}(0)
= \sum_{\alpha=1}^{L}
c_{\alpha} \mathbf{R}^{(\alpha)},
\]
with coefficients
\begin{equation}
c_{\alpha} = \sum_{n=1}^N L^{(\alpha)}_n P_n(0).
\end{equation}
The relaxation dynamics toward the stationary state reads
\begin{equation}
\mathbf{P}(t)=
\mathbf{P}^{(E)}
+ \sum_{\alpha=2}^{L} c_{\alpha} e^{\lambda_{\alpha} t} \mathbf{R}^{(\alpha)} .
\end{equation}
where we used the property $c_{1}=1$ and $\mathbf{R}^{(1)}=\mathbf{P}^{(E)}$. 
Since $\lambda_{\alpha} < 0$ for $ \alpha \ge 2$, all nonstationary components decay and
$\mathbf{P}(t)$ converges to $\mathbf{P}^{(E)}$. Let us now consider the two initial conditions $\rho_i$ and $\rho_i^{\prime}$, given by Eqs.(10) and (11) in the main text. They correspond to initial single-site excitation, namely $P_n(0)=\delta_{n,1}$, $P_n^{\prime}(0)=\delta_{n,L}$. Using Eq.(17) it then follows that 
\begin{equation}
\left| \frac{c_{\alpha}}{c^{\prime}_{\alpha}} \right|=\left| \frac{L_1^{(\alpha)}}{L_L^{(\alpha)}} \right|=\left| \frac{S_1}{S_L} \right| r^{L-1} \simeq r^{L-1}
\end{equation}
for $ \alpha \ll L$, which is Eq.(12) given in the main text

\subsection*{E. Distances for diagonal density matrices}

If $\rho(t)$ is diagonal, $\rho_{nn}(t)=P_n(t)$, the trace distance from the
stationary state is
\[
D_{\mathrm{tr}}(t)
= \frac{1}{2}
\sum_{n=1}^N
\left| P_n(t) - P^{(E)}_n \right|.
\]
while the Hilbert--Schmidt distance takes the form
\[
D_{\mathrm{HS}}(t)
=
\left[
\sum_{n=1}^N
\left( P_n(t) - P^{(E)}_n \right)^2
\right]^{1/2}.
\]

}}


\begin{thebibliography}{99}

\bibitem{r1} Mpemba E B and Osborne D G 1969 Cool? \emph{Phys. Educ.} \textbf{4} 172

\bibitem{r2} Kell G S 1969 The freezing of hot and cold water \emph{Am. J. Phys.} \textbf{37} 564

\bibitem{r3} Jeng M 2006 The Mpemba effect: when can hot water freeze faster than cold? \emph{Am. J. Phys.} \textbf{74} 514

\bibitem{r4} Bechhoefer J, Kumar A and Chetrite R 2021 A fresh understanding of the Mpemba effect \emph{Nat. Rev. Phys.} \textbf{3} 534

\bibitem{r5} Lasanta A, Vega Reyes F, Prados A and Santos A 2017 When the hotter cools more quickly: Mpemba effect in granular fluids \emph{Phys. Rev. Lett.} \textbf{119} 148001

\bibitem{r6} Lu Z and Raz O 2017 Nonequilibrium thermodynamics of the Markovian Mpemba effect and its inverse \emph{Proc. Natl. Acad. Sci. USA} \textbf{114} 5083

\bibitem{r7} Klich I, Raz O, Hirschberg O and Vucelja M 2019 Mpemba Index and Anomalous Relaxation \emph{Phys. Rev. X} \textbf{9} 021060

\bibitem{r8} Kumar A and Bechhoefer J 2020 Exponentially faster cooling in a colloidal system \emph{Nature} \textbf{584} 64

\bibitem{r9} Baity-Jesi M, Calore E, Cruz A, Maiorano A, Marinari E, Parisi G, Perez-Gaviro S, Ricci-Tersenghi F, Ruiz-Lorenzo J J, Schifano S F, Seoane B, Tripiccione R and Yllanes D 2019 The Mpemba effect in spin glasses is a persistent memory effect \emph{Proc. Natl. Acad. Sci. USA} \textbf{116} 15350

\bibitem{r9b} Biswas A, Prasad V V V, Raz O and Rajesh R 2020 Mpemba effect in driven granular Maxwell gases \emph{Phys. Rev. E} \textbf{102} 012906

\bibitem{r9c} Biswas A, Prasad V V V and Rajesh R 2023 Mpemba effect in driven granular gases: Role of distance measures \emph{Phys. Rev. E} \textbf{108} 024902

\bibitem{r10} Teza G, Yaacoby R and Raz O 2023 Relaxation Shortcuts through Boundary Coupling \emph{Phys. Rev. Lett.} \textbf{131} 017101

\bibitem{r11} Vu T V and Hayakawa H 2025 Thermomajorization Mpemba Effect \emph{Phys. Rev. Lett.} \textbf{134} 107101

\bibitem{r12} Teza G, Bechhoefer J, Lasanta A, Raz O and VuceljaI M 2025 Speedups in nonequilibrium thermal relaxation: Mpemba and related effects \emph{arXiv:2502.01758} [quant-ph]

\bibitem{r12b} Ares F, Calabrese P and Murciano S 2025 The quantum Mpemba effects \emph{Nat. Rev. Phys.} \textbf{7} 451




\bibitem{r13} Nava A and Fabrizio M 2019 Lindblad dissipative dynamics in the presence of phase coexistence \emph{Phys. Rev. B} \textbf{100} 125102

\bibitem{r14} Carollo F, Lasanta A and Lesanovsky I 2021 Exponentially Accelerated Approach to Stationarity in Markovian Open Quantum Systems through the Mpemba Effect \emph{Phys. Rev. Lett.} \textbf{127} 060401

\bibitem{r15} Manikandan S K 2021 Equidistant quenches in few-level quantum systems \emph{Phys. Rev. Res.} \textbf{3} 043108

\bibitem{r16} Kochsiek S, Carollo F and Lesanovsky I 2022 Accelerating the approach of dissipative quantum spin systems towards stationarity through global spin rotations \emph{Phys. Rev. A} \textbf{106} 012207

\bibitem{r17} Ivander F, Anto-Sztrikacs N and Segal D 2023 Hyperacceleration of quantum thermalization dynamics by bypassing long-lived coherences: An analytical treatment \emph{Phys. Rev. E} \textbf{108} 014130

\bibitem{r18} Dong J W, Zhang S, Huang Y, Li X, Zhang X and Chen Y 2025 Quantum Mpemba effect of localization in the dissipative mosaic model \emph{Phys. Rev. A} \textbf{111} 022215

\bibitem{r19} Zhou Y-L, Zhang X, Gao Y and Guo G-C 2023 Accelerating relaxation through
Liouvillian exceptional point  \emph{Phys. Rev. Res.} \textbf{5} 043036

\bibitem{r20} Chatterjee A K, Takada S and Hayakawa H 2023 Quantum Mpemba effect in a quantum dot with reservoirs \emph{Phys. Rev. Lett.} \textbf{131} 080402

\bibitem{r21} Moroder M, Dabelow L, Engel A and Esposito M 2024 Thermodynamics of the quantum Mpemba effect \emph{Phys. Rev. Lett.} \textbf{133} 140404

\bibitem{r22} Chatterjee A K, Takada S and Hayakawa H 2024 Multiple quantum Mpemba effect: exceptional points and oscillations \emph{Phys. Rev. A} \textbf{110} 022213

\bibitem{r28} Nava A and Egger R 2024 Mpemba Effects in Open Nonequilibrium Quantum Systems \emph{Phys. Rev. Lett.} \textbf{133} 136302

\bibitem{r29} Shapira S A, Nava A, Egger R and Giuliano D 2024 Inverse Mpemba effect demonstrated
on a single trapped ion qubit \emph{Phys. Rev. Lett.} \textbf{133} 010403

\bibitem{r29b}
Shapira S A, Shapira Y, Markov J, Teza G, Akerman N, Raz O and Ozeri R 2024 Inverse Mpemba effect demonstrated on a single trapped ion qubit \emph{Phys. Rev. Lett.} \textbf{133} 010403


\bibitem{r30} Wang X and Wang J 2024 Mpemba effects in nonequilibrium open quantum system \emph{Phys. Rev. Res.} \textbf{6} 033330

\bibitem{r31} Liu D, He Y, Wang S and Chen Y 2024 Speeding up quantum heat engines by the
Mpemba effect  \emph{Phys. Rev. A} \textbf{110} 042218

\bibitem{r34} Longhi S 2024 Bosonic Mpemba effect with non-classical states of light \emph{APL Quantum} \textbf{1} 046110

\bibitem{r35} Longhi S 2025 Laser Mpemba effect \emph{Opt. Lett.} \textbf{50} 2069

\bibitem{r35b} Longhi S 2024 Photonic Mpemba effect \emph{Opt. Lett.} \textbf{49} 5188

\bibitem{r36} Longhi S 2025 Mpemba effect and super-accelerated thermalization in the damped quantum harmonic oscillator \emph{Quantum} \textbf{9} 1677

\bibitem{r37} Dong J W, Zhang X, Li X and Chen Y 2025 Quantum Mpemba effect of localization in the dissipative mosaic model \emph{Phys. Rev. A} \textbf{111} 022215

\bibitem{r38} Zhang J, Liu Z, Chen X and Zhao Y 2025 Observation of quantum strong Mpemba effect \emph{Nat. Commun.} \textbf{16} 301

\bibitem{r40} Furtado J and Santos A C 2025 Enhanced quantum Mpemba effect with squeezed thermal reservoirs \emph{Ann. Phys.} \textbf{480} 170135

\bibitem{r42} Qian D, Wang H and Wang J 2025 Intrinsic quantum Mpemba effect in Markovian systems and quantum circuits \emph{Phys. Rev. B} \textbf{111} L220304

\bibitem{r43} Longhi S 2025 Quantum Mpemba effect from initial system-reservoir entanglement \emph{APL Quantum} \textbf{2} 026133

\bibitem{r44} Bao R and Hou Z 2025 Accelerating Quantum Relaxation via Temporary Reset: A Mpemba-Inspired Approach
\emph {Phys. Rev. Lett.} \textbf{ 135} 150403 

\bibitem{r45}  Strachan D J, Purkayastha A and Clark S R 2025 Non-Markovian quantum Mpemba effect \emph{Phys. Rev. Lett.} \textbf{134} 220403

\bibitem{r45b} Longhi S 2025 Quantum Mpemba effect from non-normal dynamics \emph{Entropy} \textbf{27} 581

\bibitem{r50}  Medina I, Culhane O, Binder F C, Landi G T and Goold J 2025 Anomalous discharging of quantum batteries: the ergotropic Mpemba effect \emph{Phys. Rev. Lett.} \textbf{134} 220402

\bibitem{r51} Bagui P, Chatterjee A and Agarwalla B K 2025 Accelerated relaxation and Mpemba-like effect for operators in open quantum systems \emph{arXiv:2510.24630} [quant-ph]


\bibitem{r52} Mondal S and Sen U 2025 Mpemba effect in self-contained quantum refrigerators: accelerated cooling \emph{arXiv:2507.15811} [quant-ph]

\bibitem{r52b}
Xu M, Wei Z, Jiang X-P and Pan L 2025 Expedited thermalization dynamics in incommensurate systems \emph{arXiv:2505.03645} [quant-ph]

\bibitem{r53} 
Ma W and Liu J 2025 Quantum Mpemba effect in parity-time symmetric systems \emph{arXiv:2508.17575} [quant-ph]

\bibitem{r54}  
Wei Z, Xu M, Jiang X-P, Hu H and Pan L 2025 Quantum Mpemba effect in dissipative spin chains at criticality \emph{arXiv:2508.18906} [quant-ph]









\bibitem{r18a} 
Ares F, Murciano S and Calabrese P 2023 Entanglement asymmetry as a probe of symmetry breaking \emph{Nat. Commun.} \textbf{14} 2036

\bibitem{r23} 
Joshi L K, Franke J, Rath A, Ares F, Murciano S, Kranzl F, Blatt R, Zoller P, Vermersch B, Calabrese P, Roos C F and Joshi M K 2024 Observing the quantum Mpemba effect in quantum simulations \emph{Phys. Rev. Lett.} \textbf{133} 010402

\bibitem{r24} 
Rylands C, Klobas K, Ares F, Calabrese P, Murciano S and Bertini B 2024 Microscopic origin of the quantum Mpemba effect in integrable systems \emph{Phys. Rev. Lett.} \textbf{133} 010401

\bibitem{r25} 
Murciano S, Ares F, Klich I and Calabrese P 2024 Entanglement asymmetry and quantum Mpemba effect in the XY spin chain \emph{J. Stat. Mech.} 013103

\bibitem{r32} 
Caceffo F, Murciano S and Alba V 2024 Entangled multiplets, asymmetry, and quantum Mpemba effect in dissipative systems \emph{J. Stat. Mech.} 063103

\bibitem{r33} 
Liu S, Zhang H-K, Yin S and Zhang S-X 2024 Symmetry restoration and quantum Mpemba effect in symmetric random circuits \emph{Phys. Rev. Lett.} \textbf{133} 140405

\bibitem{r26} 
Chalas K, Ares F, Rylands C and Calabrese P 2024 Multiple crossing during dynamical symmetry restoration and implications for the quantum Mpemba effect \emph{J. Stat. Mech.} 103101

\bibitem{r27} 
Yamashika S, Ares F and Calabrese P 2024 Entanglement asymmetry and quantum Mpemba effect in two-dimensional free-fermion systems \emph{Phys. Rev. B} \textbf{110} 085126

\bibitem{r39} 
Yamashika S, Calabrese P and Ares F 2024 Quenching from superfluid to free bosons in two dimensions: entanglement, symmetries, and quantum Mpemba effect \emph{arXiv:2410.14299}

\bibitem{r41} 
Liu S, Zhang H-K, Yin S, Zhang S-X and Yao H 2024 Quantum Mpemba effects in many-body localization systems \emph{arXiv:2408.07750}




\bibitem{P1} 
Nava A and Egger R 2025 Pontus-Mpemba effects \emph{Phys. Rev. Lett.} \textbf{135} 140404

\bibitem{P2} 
Yu H, Hu J and Zhang S-X 2025 Quantum Pontus-Mpemba effects in real and imaginary-time dynamics \emph{arXiv:2509.01960} [quant-ph]


\bibitem{P3} 
Nava A, Egger R, Dey B and Giuliano D 2025 Speeding up Pontus-Mpemba effects via dynamical phase transitions \emph{arXiv:2509.09366} [quant-ph]

\bibitem{P4} 
Aditya S, Summer A, Sierant P and Turkeshi X 2025 Mpemba effects in quantum complexity \emph{arXiv:2509.22176} [quant-ph]




\bibitem{Ashida2020} Ashida Y, Gong Z and Ueda M 2020 Non-Hermitian physics \emph{Adv. Phys.} \textbf{69} 249




\bibitem{NH1} Yao S and Wang Z 2019 Edge states and topological invariants of non-Hermitian systems \emph{Phys. Rev. Lett.} \textbf{121} 086803
\bibitem{NH2} Lee C H and Thomale R 2019 Anatomy of skin modes and topology in non-Hermitian systems \emph{Phys. Rev. B} \textbf{99} 201103
\bibitem{NH3} Yokomizo K and Murakami S 2019 Non-Bloch band theory of non-Hermitian systems \emph{Phys. Rev. Lett.} \textbf{123} 066404
\bibitem{NH4} Kunst F K, Edvardsson E, Budich J C and Bergholtz E J 2018 Biorthogonal bulk?boundary correspondence in non-Hermitian systems \emph{Phys. Rev. Lett.} \textbf{121} 026808
\bibitem{NH5} Longhi S 2019 Probing non-Hermitian skin effect and non-Bloch phase transitions \emph{Phys. Rev. Research} \textbf{1} 023013
\bibitem{NH6} Foa Torres L E F 2019 Perspective on topological states of non-Hermitian lattices \emph{J. Phys. Mater.} \textbf{3} 014002
\bibitem{NH7} Song F, Yao S and Wang Z 2019 Non-Hermitian skin effect and chiral damping in open quantum systems \emph{Phys. Rev. Lett.} \textbf{123} 170401
\bibitem{NH8} Longhi S 2020 Unraveling the non-Hermitian skin effect in dissipative systems \emph{Phys. Rev. B} \textbf{102} 201103(R)
\bibitem{NH9} Borgnia D S, Kruchkov A J and Slager R J 2020 Non-Hermitian boundary modes and topology \emph{Phys. Rev. Lett.} \textbf{124} 056802
\bibitem{NH10} Okuma N, Kawabata K, Shiozaki K and Sato M 2020 Topological origin of non-Hermitian skin effects \emph{Phys. Rev. Lett.} \textbf{124} 086801
\bibitem{NH11} Zhang K, Yang Z and Fang C 2020 Correspondence between winding numbers and skin modes in non-Hermitian systems \emph{Phys. Rev. Lett.} \textbf{125} 126402
\bibitem{NH12} Li L, Lee C H, Mu S and Gong J 2020 Critical non-Hermitian skin effect \emph{Nat. Commun.} \textbf{11} 5491
\bibitem{NH13} Xiao L \emph{et al.} 2020 Non-Hermitian bulk-boundary correspondence in quantum dynamics \emph{Nat. Phys.} \textbf{16} 761
\bibitem{NH14} Yang Z, Zhang K, Fang C and Hu J 2020 Non-Hermitian bulk-boundary correspondence and auxiliary generalized Brillouin zone theory \emph{Phys. Rev. Lett.} \textbf{125} 226402
\bibitem{NH15} Ghatak A, Brandenbourger M, van Wezel J and Coulais C 2020 Observation of non-Hermitian topology and its bulk-edge correspondence in an active mechanical metamaterial \emph{Proc. Natl. Acad. Sci. USA} \textbf{117} 29561
\bibitem{NH15b} Longhi S 2025 Erratic non-Hermitian Skin Localization \emph{Phys. Rev. Lett.} \textbf{134} 196302
\bibitem{NH16} Bergholtz E J, Budich J C and Kunst F K 2021 Exceptional topology of non-Hermitian systems \emph{Rev. Mod. Phys.} \textbf{93} 015005
\bibitem{NH17} Zhang K, Yang Z and Fang C 2022 Universal non-Hermitian skin effect in two and higher dimensions \emph{Nat. Commun.} \textbf{13} 2496
\bibitem{NH18} Ding K, Fang C and Ma G 2022 Non-Hermitian topology and exceptional-point geometries \emph{Nat. Rev. Phys.} \textbf{4} 745
\bibitem{NH19} Zhang X, Zhang T, Lu M H and Chen Y F 2022 A review on non-Hermitian skin effect \emph{Adv. Phys.: X} \textbf{7} 2109431
\bibitem{NH20} Okuma N and Sato M 2023 Non-Hermitian topological phenomena: a review \emph{Annu. Rev. Condens. Matter Phys.} \textbf{14} 83
\bibitem{NH21} Lin R, Tai T, Li L and Lee C H 2023 Topological non-Hermitian skin effect \emph{Front. Phys.} \textbf{18} 53605
\bibitem{NH22} Banerjee A, Sarkar R, Dey S and Narayan A 2023 Non-Hermitian topological phases: principles and prospects \emph{J. Phys.: Condens. Matter} \textbf{35} 333001
\bibitem{NH23} Gohsrich J T, Banerjee A and Kunst F K 2025 The non-Hermitian skin effect: a perspective \emph{EPL} \textbf{150} 60001



\bibitem{LS1} 
Haga T, Nakagawa M, Hamazaki R and Ueda M 2021 Liouvillian skin effect: slowing down of relaxation processes without gap closing \emph{Phys. Rev. Lett.} \textbf{127} 070402
\bibitem{LS2}
 Yang F, Jiang Q-D and Bergholtz E J 2022 Liouvillian skin effect in an exactly solvable model \emph{Phys. Rev. Research} \textbf{4} 023160
\bibitem{LS3}
 Sannia A, Giorgi G L, Longhi S and Zambrini R 2025 Liouvillian skin effect in quantum neural networks \emph{Optica Quantum} \textbf{3} 189-194
\bibitem{LS4}
Cai D-H, Yi W and Dong C-X 2025 Optical pumping through the Liouvillian skin effect \emph{Phys. Rev. B} \textbf{111} L060301
\bibitem{LS5}
 Mao L, Yang X, Tao M-J, Hu H and Pan L 2024 Liouvillian skin effect in a one-dimensional open many-body quantum system with generalized boundary conditions \emph{Phys. Rev. B} \textbf{110} 045440
\bibitem{LS6}
 Shigedomi Y and Yoshida T 2025 Liouvillian skin effects in two-dimensional electron systems at finite temperatures \emph{arXiv:2505.18001} [cond-mat.mes-hall]
 
 
\bibitem{H1}
Eisler V 2011 Crossover between ballistic and diffusive transport: The quantum exclusion process \emph{J. Stat. Mech.} \textbf{2011} P06007

\bibitem{H2}
Temme K, Wolf M M and Verstraete F 2012 Stochastic exclusion processes versus coherent transport \emph{New J. Phys.} \textbf{14} 075004

\bibitem{H3}
Solanki P, Cabot A, Brunelli M, Carollo F, Bruder C and Lesanovsky I 2025 Generation of entanglement and nonstationary states via competing coherent and incoherent bosonic hopping \emph{Phys. Rev. A} \textbf{112} L030601



\bibitem{IMP1}
Schreiber A, Cassemiro K N, Potccek V, Gabris A, Jex I and Silberhorn C 2011 Decoherence and disorder in quantum walks: From ballistic spread to localization \emph{Phys. Rev. Lett.} \textbf{106} 180403

\bibitem{IMP2}
Lin Q, Yi W and Xue P 2023 Manipulating directional flow in a two-dimensional photonic quantum walk under a synthetic magnetic field \emph{Nat. Commun.} \textbf{14} 6283

\bibitem{IMP3}
Longhi S 2024 Incoherent non-Hermitian skin effect in photonic quantum walks \emph{Light Sci. Appl.} \textbf{13} 95

\bibitem{IMP4}
Garbe L, Minoguchi Y, Huber J and Rabl P 2024 The bosonic skin effect: Boundary condensation in asymmetric transport \emph{SciPost Phys.} \textbf{16} 029






\end{thebibliography}
\end{document}